\documentclass[prl,aps,twocolumn,floatfix,showpacs,superscriptaddress,unsortedaddress]{revtex4}
\usepackage{graphicx}
\usepackage{amssymb}
\usepackage{bm}
\usepackage{psfrag}

\begin{document}
\title{Magnetic field enhanced coherence length in cold atomic gases}
\author{O. Sigwarth}
\email{sigwarth@spectro.jussieu.fr}
\affiliation{Laboratoire Kastler Brossel, Tour 12, Etage 1, 4 Place Jussieu, F-75005 Paris, France}
\author{G. Labeyrie}
\affiliation{Institut Non Lin\'eaire de Nice, 1361 route des Lucioles, F-06560 Valbonne, France.}
\author{T. Jonckheere}
\altaffiliation[Now at: ]{Centre de Physique Th\'eorique, Campus de Luminy,
case 907, 13009 Marseille, France.}
\affiliation{Laboratoire Kastler Brossel, Tour 12, Etage 1, 4 Place Jussieu, F-75005 Paris, France}
\author{D. Delande}
\affiliation{Laboratoire Kastler Brossel, Tour 12, Etage 1, 4 Place Jussieu, F-75005 Paris, France}
\author{R. Kaiser}
\affiliation{Institut Non Lin\'eaire de Nice, 1361 route des Lucioles, F-06560 Valbonne, France.}
\author{C. Miniatura}
\affiliation{Institut Non Lin\'eaire de Nice, 1361 route des Lucioles, F-06560 Valbonne, France.}

\date{\today}

\begin{abstract}
We study the effect of an external magnetic field on coherent
backscattering (CBS) of light from a cold rubidium vapor. We
observe that the CBS enhancement factor can be {\it increased}
with $B.$ This surprising behavior shows that the coherence length
of the system can be increased by adding a magnetic field, in
sharp contrast with usual situations. This is mainly due to the
lifting of the degeneracy between Zeeman sublevels. We find good
agreement between our experimental data and a full Monte-Carlo
simulation, taking into account the magneto-optical effects and
the geometry of the atomic cloud.
\end{abstract}


\maketitle

During the last few years, ultracold atomic gases have become an ideal experimental and theoretical tool to
investigate quantum phase transitions. Well-known examples are the
Bose-Einstein condensation, degenerate Fermi gases and the
Mott-Hubbard metal-insulator transition~\cite{Greiner}. Cold
atomic gases also constitute extremely interesting samples to
study weak and strong localization of waves in disordered media.
Indeed, contrary to typical condensed matter samples (e.g.
multiple scattering of electrons in metals or semiconductors),
atoms interact very resonantly with light. Thus the scattering
processes at work can be more finely controlled and analyzed,
opening the way to new studies in localization phenomena. Of
special interest is the so-called mesoscopic regime, where
interefences between multiply scattered waves are not washed out
by the disorder. Here the notion of coherence length $L_{\phi}$ is
crucial since it sets the characteristic length scale at which
interference effects play a role in transport. For sample sizes $L
<L_{\phi}$, these interference effects give rise to the so-called
weak localization corrections of transport \cite{Houches}. They
originate from the interference between waves contrapropagating
along loops in a scattering path. These interferences survive the
configuration average and generally induce a negative correction
to the usual classical Drude-Boltzmann conductance, as evidenced
in 2D-electron gases~\cite{Bergmann}. In optics, a related
phenomenon is the coherent backscattering (CBS) effect~\cite{CBS}.
Here the diffuse reflexion off the sample is enhanced around
backscattering in a narrow angular range $\propto 1/(k\ell)\ll 1$
($k$ is the light wavevector and $\ell$ the scattering mean free
path) because of constructive interferences between waves
propagating along reversed paths. This interference effect is
similar to the previous closed-loops interference and is maximum
when the complex amplitudes of the reversed paths are exactly
balanced. This is the case for time-reversal symmetric systems
where, for a convenient choice of polarization channel, an
enhancement factor close to 2 is observed~\cite{CBS}.

These interferential corrections to transport are particularly
sensitive to any physical source of dephasing susceptible to
suppress interference. Some examples are inelastic scattering
(electron-phonon interaction \cite{Houches}, saturation of an
atomic transition \cite{TC}), spin-flip scattering \cite{Houches}
or internal degeneracies \cite{StructInt}. In this respect, the
use of a magnetic field, which breaks time-reversal invariance,
offers an efficient tool to investigate phase coherent effects
\cite{Bergmann}. In an electron gas, the loops of a diffusion path
now enclose a magnetic flux which alters interference effects. As
a consequence the coherence length is reduced as $B$ is increased.
These modifications of the weak localization corrections have been
experimentally verified on magnetoconductance measurements
\cite{Bergmann}. In optics, one gets the same kind of results but
evidenced in CBS studies. Adding an external magnetic field breaks
the time-reversal symmetry, leading to unbalanced interfering
amplitudes and consequently to a {\em decrease} of the enhancement
factor~\cite{CBSBth,CBSBexp}. This is so because, by modifying the
refractive index of the medium, the magnetic field induces a
Faraday effect which alters the phase difference at propagation
and, in turn, gradually destroys the CBS effect.

In this letter, we report the first study of the CBS enhancement
factor with cold atoms subjected to an external magnetic field
$B$. We show an \emph{a priori} surprising {\em restoration} of
interference effects. Correspondingly, the coherence length is
\textit{increased}. This rather paradoxical result stems from the
existence of an internal structure of the atoms (a non-zero
angular momentum, hence degenerate, ground state) which is already
responsible, at zero field, for unbalanced interfering amplitudes
and reduced interference contrast~\cite{StructInt,Havey,mc}. The
magnetic field breaks the degeneracy of the atomic ground state
and splits the atomic resonance line (Zeeman effect). As $B$ is
increased, less and less Zeeman levels participate in the
scattering process. At large enough fields, atoms behave as
effective non-degenerate two-level systems and full multiple
scattering interference contrast is restored.

The paper is organized as follows. We first briefly describe the
experiment and our observations. We then discuss the basic
physical ingredients and the principle of the Monte-Carlo
simulation. Then, we identify the various mechanisms at work,
trying to give some simple physical pictures.

\begin{figure}
\begin{center}
\psfrag{k}{${\mathbf k}_i$}
\includegraphics[width=7cm]{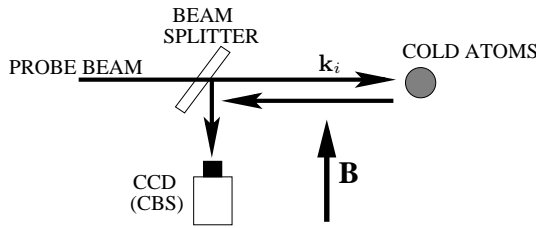}
\end{center}
\caption{\label{fig1} Principle of the experiment. An external
magnetic field ${\mathbf B}$ is applied orthogonally to the
incident wavevector ${\mathbf k}_i.$ A circularly-polarized laser
beam is shined at the cold atomic cloud, and the backscattered
far-field intensity distribution is recorded on a CCD (detection
in the parallel helicity channel).}
\end{figure}

\begin{figure}
\begin{center}
\includegraphics[width=8cm,height=6cm]{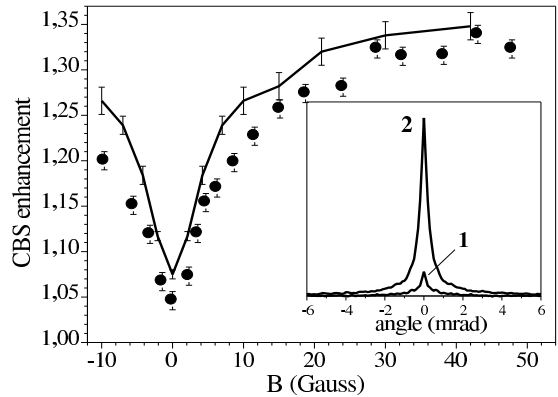}
\end{center}
\caption{\label{alpha} CBS enhancement factor of light on a cold
Rb atomic cloud, measured in the parallel helicity channel
$h\,{\parallel}\, h$, as a function of the transverse magnetic
field $B$ (circles); it shows a dramatic {\em increase} at
non-zero magnetic field whereas time-reversal symmetry is broken.
For each $B$ value, the laser is tuned to resonance with the
$m_{F}=3 \rightarrow m_{F^{\prime}}=4$ transition. The solid line
is the result of the Monte-Carlo simulation (no adjustable
parameter). The inset shows two (angularly averaged) experimental
CBS peak profiles for $B$=0 (1) and $B$=43 G (2).}
\end{figure}

Our CBS experimental setup and its typical relevant parameters
have been described in~\cite{StructInt}. We have here added a pair
of Helmholtz coils to produce a uniform external magnetic field
$\textbf{B}$ perpendicular to the incoming light wavevector
$\mathbf{k}_i$ (Fig.\ref{fig1}). The CBS probe is a weakly
saturating laser beam (linewidth less than 2MHz) with initial
circular polarization \bm{$\epsilon_i$}, shined at the atomic
cloud while the magneto-optical trap (MOT) is off. It is
quasi-resonant with the closed $F=3\rightarrow F^{\prime }=4$
transition of the D2 line of Rb$^{85}$ (wavelength $\lambda =780$
nm, linewidth $\Gamma/2\pi=5.9$ MHz). The magnetic field is
switched on slightly before the CBS probe is applied and the
optical thickness at zero field is $b=31$. During the
measurements, for each value of $B$, the CBS laser is
\emph{maintained at resonance with the Zeeman shifted
$m_{F}=3\rightarrow m_{F'}=4$ transition} (quantization axis along
$\mathbf{B}$). This is an essential feature to understand our
results and the interpretation given below. The CBS signal is
recorded in the so-called parallel helicity channel, where the
detected helicity is identical to the incident one. Results are
shown in Fig.\ref{alpha}, where the CBS enhancement factor
$\alpha$ (peak to background ratio) is plotted as a function of
$B$. It is observed that $\alpha$, initially very small ($\approx
1.05$ for $B$=0), sharply increases with $B$ and saturates around
1.33 above 30G. This behavior is obviously very different from the
classical case, where $\alpha$ is reduced by the presence of $B$
in all reported instances \cite{CBSBexp}.

In the weak localization regime~\cite{AdBart}, a multiple
scattering path is depicted as a succession of single scattering
events (described by their scattering amplitudes) separated by
average propagation in the effective medium (described by the
refractive optical index). When a magnetic field $\mathbf{B}$ is
applied, the atomic energy levels are shifted by an amount $\Delta
E = gm\mu B\hbar$, where $g$ is the Land\'e factor (respectively
1/3 and 1/2 for the ground and excited state considered here), $m$
the magnetic quantum number and $\mu/2\pi = 1.4$ MHz/G (Bohr
magneton). The Zeeman splitting into a non-degenerate structure
affects both the properties of an individual scatterer and of the
effective medium \cite{Hanle}. An essential difference between
usual scatterers and atoms is that atoms have an extremely
resonant scattering cross-section (quality factor $\simeq 10^8$).
Hence few Gauss are enough to induce a Zeeman shift comparable to
the resonance linewidth $\Gamma$, and consequently a large
modification of the scattering cross-section. This is also why the
CBS enhancement factor displays dramatic changes, even at moderate
$B,$ in sharp contrast with published experimental results on
standard scatterers which required magnetic fields in the Tesla
range \cite{CBSBexp}. An additional complication with atoms is the
existence of several Zeeman sublevels. The scattering process can
induce a transition between different Zeeman sublevels (inelastic
Raman scattering).

The scattering amplitude of an incident photon
$|\mathbf{k}\bm{\epsilon}\rangle$ onto the outgoing light state
$|\mathbf{k^{\prime}}\bm{\epsilon^{\prime}}\rangle$ by an atom
whose initial state $|Fm_1\rangle$ is changed to $|Fm_2\rangle$ is
described by the scattering tensor $T_{\!m_{1},m_{2}}$:
 \begin{equation}\label{scat}
\overline{\bm{\epsilon'}}.T_{\!m_{1},m_{2}}.\bm{\epsilon}
  =\!\!\!\!\!\!
 \sum_{m'=-F'}^{F'}\!\!\!\!\!\frac{\langle Fm_{2}|
 {\mathbf d}.\overline{\bm{\epsilon'}}|
 F'm'\rangle\langle F'm'|
 {\mathbf d}.\bm{\epsilon}|Fm_{1}\rangle}{\delta+\mu B(gm_{1}-
 g'm')+i\Gamma/2}
 \end{equation}
where ${\mathbf d}$ is the atomic dipole operator and $\delta=\omega-\omega_{0}$
is the detuning of the laser frequency $\omega$ from the zero-field atomic
 frequency $\omega_{0}$.
The scattered photon
has a frequency $\omega'=\omega+g\mu B(m_1-m_2)$
and polarization
 $\Delta_{\mathbf k'}T_{m_{1},m_{2}}\bm{\epsilon}$, where $\Delta_{\mathbf k'}$
 is the projector on the plane perpendicular to  ${\mathbf k'}$. In eq.~(\ref{scat}), the
 denominator is simply the detuning of the incoming laser from the
 (Zeeman shifted) $|Fm_1\rangle \to |F'm'\rangle$ transition.

In the presence of a magnetic field, the effective medium displays well documented
magneto-optical anisotropies
\cite{VoigtCold, CBSBth}: a polarization propagating along a given
direction $\hat{\textbf{r}}$ decomposes on two eigenmodes which propagate with a
differential phaseshift (Faraday effect) and a differential
attenuation (Cotton-Mouton effect). When atoms are uniformly distributed
over the Zeeman ground states (a good
approximation for atoms produced in a MOT) and when
optical pumping is negligible \cite{pompage}, the
complex refractive index matrix reads:
 \begin{equation}
 \label{index}
 N(\omega,\hat{\textbf{r}}) = 1-\frac{1}{2k\ell_0} \ \frac{\Gamma/2}{\delta+\mathrm{i}\Gamma/2} \Delta_{\bm r}\left(\begin{array}{ccc}
 \zeta& \eta&0\\
 -\eta&\zeta&0\\
 0&0&\zeta+\xi  \end{array} \right)\Delta_{\bm r}
 \end{equation}
Here $\ell_0$ is the mean free path at zero field, $\Delta_{\bm
r}$ is the projector onto the plane transverse to
$\hat{\textbf{r}}$ and $\zeta$, $\eta$ and $\xi$ are simple
functions of $\phi=\frac{i\mu B}{\delta+i\Gamma/2}$ ($\zeta$ and
$\xi$ are even in $B$ whereas $\eta$ is odd). Their detailed
expressions will be published elsewhere. $\zeta$ describes the
normal extinction of the light in the medium, while $\eta$ and
$\xi$ respectively describe the Faraday rotation and the
Cotton-Mouton effect.

\begin{figure}
\psfrag{Icoh/Iinc}{$I_{\mathrm{coh}}/I_{\mathrm{inc}}$}
\begin{center}
\includegraphics[width=8cm,angle=0]{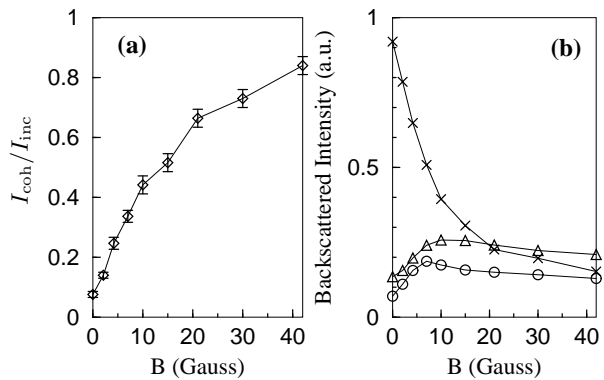}
\end{center}
\caption{\label{fig4} Results of the Monte-Carlo simulations. (a)
shows the increase of the ratio of coherent over incoherent
intensity with $B$. (b) Contribution of single scattering
(triangles), incoherent (crosses) and coherent (circles) multiple
scattering to the total intensity in the backward direction. Lines
are drawn to guide the eyes.}
\end{figure}

The scattering~(\ref{scat}) and the index~(\ref{index})
matrices are the basic ingredients used in a Monte-Carlo
simulation \cite{mc}. The enhancement factor at exact
backscattering is obtained as
 $\alpha=1+I_{\mathrm{coh}}/(I_{\mathrm{single}}+I_{\mathrm{inc}})$ where
$I_{\mathrm{single}}$ is the single scattering contribution,
$I_{\mathrm{inc}}$ the diffuse contribution (incoherent intensity)
and $I_{\mathrm{coh}}$ the multiple scattering interference
contribution (coherent intensity). Our numerical accuracy limits
calculations up to $B \approx 40 G$. Results are shown in
Figs.~\ref{alpha} and \ref{fig4}. The calculated enhancement
factor is rapidly increasing from $\alpha=1.07$ at 0 G to
$\alpha=1.35$
 at 42 G, in good quantitative agreement with the experimental observation. This agreement
proves that our approach catches the most important physical
mechanisms at work. In order to identify these mechanisms, we plot
in Fig.~\ref{fig4}(a), the ratio of the coherent over incoherent
signals as a function of $B$. At $B=0$, the coherent intensity is
much smaller than the incoherent one, featuring the CBS scrambling
due to the atomic internal structure \cite{StructInt}. Then
$I_{\mathrm{coh}}/I_{\mathrm{inc}}$ increases with $B$, featuring
the restoration of the interference. As can be seen in
Fig.~\ref{fig4}(b), this is mainly due to the strong decrease of
the incoherent intensity (crosses) even if the coherent intensity
(circles) displays a small increase at small $B$. This decrease is
related to the decrease of the atomic cross section: the optical
thickness is lowered and multiple scattering is reduced. On the
contrary, the coherent intensity increases at small $B$, because
the reduction of the atomic degeneracy restores the interference.
This phenomenon competes with the reduction of multiple
scattering: at large $B$, the coherent intensity decreases, but
more slowly than the incoherent one. Note, however, that the CBS
enhancement factor remains noticeably smaller than 2, even at
large $B$, due to single scattering. The relative importance of single
scattering is due to our specific cloud geometry \cite{mc}.

At large magnetic field $\mu B \gg \Gamma$, the Zeeman structure
is sufficiently split and the $m_{F}=3\rightarrow m_{F'}=4$
optical transition remains the only one to be excited since the
incoming light has been set at resonance with it ($\delta = \mu
B$). $I_{\mathrm{coh}}/I_{\mathrm{inc}}$ then approaches 1 because
one can envision this transition as being closed and isolated,
\emph{yielding an effective two-level system with no internal
degeneracy}. As a consequence, only left-circular photons with
respect to \textbf{B} can be absorbed and emitted at each
scattering and full interference contrast is restored.  Note that the single scattering contribution cannot
be removed by polarization analysis without removing the multiple
scattering contribution. We have
also checked that the limit $I_{\mathrm{coh}}/I_{\mathrm{inc}} \to
1$ is achieved independently of the polarization channel and of
the relative orientation of $\textbf{B}$ and $\textbf{k}_i$.

\begin{figure}
\psfrag{lphi/l}{$p_0=L_{\phi}/\ell$}
\begin{center}
\includegraphics[width=7cm,angle=0]{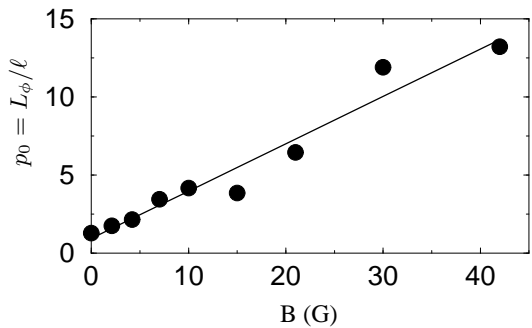}
\end{center}
\caption{\label{Lcoh} Plot of $p_0 = L_{\phi}/\ell$ \emph{versus}
$B$, as extracted from our Monte-Carlo calculation, as a function
of the magnetic field $B.$ In sharp contrast with the usual
behaviour, $p_0$ {\emph increases} with $B,$ roughly linearly (the
solid line is drawn to guide the eye). This is due to the lifting
of the Zeeman degeneracy. The points dispersion reflects the
numerical accuracy. }
\end{figure}

This restoration of the multiple scattering interference contrast
can be associated with an \emph{increase} of the coherence length
$L_{\phi}$ as $B$ is increased. Indeed, the contributions of the
various scattering orders $p$ can be extracted from the
Monte-Carlo calculation. For each value of $B$, we found a roughly
exponential decrease of the ratio
$I^{(p)}_{\mathrm{coh}}/I^{(p)}_{\mathrm{inc}} \approx
\exp(-p/p_0)$. The coherence length is simply $L_{\phi} = p_0 \,
\ell$ where $\ell$ is the average scattering mean free path
calculated at $B$. As reported in Fig.~\ref{Lcoh}, $p_0$
\textit{increases} roughly linearly with $B$. This is in sharp
contrast with usual solid-state samples where $L_{\phi}$ is
\textit{reduced} because of the Faraday effect which is
proportional to the path length and to the magnetic field. In our
case, the Faraday effect can be shown to vanish at large $B$
while, at the same time, the destructive impact of the Zeeman
degeneracy is removed, as previously discussed. The residual
coherence loss is thus due to deviations from the two-level
approximation, i.e. residual scattering amplitude due to atomic
transitions involving other (detuned from resonance) Zeeman
sublevels. From eq.~(\ref{scat}), this amplitude scales as $1/B,$
yielding a coherence length proportional to $B,$ in agreement with
our numerical results. The mesoscopic regime is achieved when
$L_{\phi} \approx L$, or equivalently when $p_0 \approx b$. At
large $B$, the optical thickness is $b \approx 5$, hence the
mesoscopic criterion is fulfilled when $B \gtrsim 15 G$.

Using our Monte-Carlo calculation, we have also quantitatively
studied the impact of magneto-optical effects at propagation
(Eq.\ref{index}) on the enhancement factor. They also appear to be
negligible for $\mu B<\Gamma$ ($\mu B=\Gamma$ corresponds to
$B=4.2$ G in our experiment). For intermediate $B$, they cause the
enhancement factor to be slightly lowered. In any case, they play
no crucial role in the increase of $\alpha$ and can be neglected
in a qualitative approach. This situation is very different from
the classical one, where Faraday rotation is the keystone to
explain the height and shape of the CBS cone \cite{CBSBexp}.

This means, in turn, that the physics at work here is the drastic
modification of the scattering properties induced by the effective
reduction of the internal structure. This effect has to be
compared with the one observed in normal metal rings seeded with
paramagnetic impurities \cite{Washburn}. There, at low fields, the
Aharonov-Bohm (AB) oscillation amplitude is strongly reduced
because of spin-flip scattering. This is equivalent to the
observation of low CBS enhancement factors because of internal
degeneracies. At high fields and low temperatures ($\mu B \gg
k_BT$), all impurities and electrons spins are aligned with
\textbf{B}. As in our experiment, $\mathbf{B}$ freezes the
internal degrees of freedom of the scatterers. Electron spin
fluctuations at scattering are then suppressed and full
AB-oscillation amplitude is recovered.

 In conclusion, we have observed a large increase of the CBS enhancement
factor in the presence of a magnetic field,
 which splits the degenerate atomic transition
into several Zeeman components.
At high magnetic field, most of them are out of resonance, leaving an effective
 two-level system where
the two interfering paths have almost equal amplitudes.
Use of this
magnetically-induced effective reduction of the atomic internal
degeneracy may offer an interesting route in the quest of strong
localization of light in cold atomic vapors.

We thank CNRS and the\ PACA Region for financial
support. Laboratoire Kastler Brossel is laboratoire
de l'Universit{\'e} Pierre et Marie
Curie et de l'Ecole Normale Sup{\'e}rieure, UMR 8552 du CNRS.


\begin{thebibliography}{99}
\bibitem{pompage} This assumption is likely to be correct if the incident light is
kept resonant with the $m_F=3\rightarrow m_{F'}=4$ transition while increasing $B$,
like in our experiment.
\end{thebibliography}
\end{document}